\documentstyle[epsfig]{mn}

\title[]{Prediction on the very early Afterglow of X-ray Flashes}

\author[]{Y. Z. Fan$^{1,2 \star}$, D. M. Wei$^{1,2 \star}$ and C. F.
Wang$^{1,2}$
\thanks{E-mail: yzfan@pmo.ac.cn(YZF); dmwei@pmo.ac.cn(DMW); amethyst@pmo.ac.cn(CFW)} \\
$^1${\sl Purple Mountain Observatory, Chinese Academy of
Science, Nanjing 210008, China}\\
       $^2${\sl National Astronomical
Observatories, Chinese Academy of Sciences, Beijing 100012,
China}}

\date{Accepted ......  Received ......; in original form 2004 March 25}
\begin{document}

\maketitle
\begin{abstract}
In the past two years, tremendous progresses to understand X-ray Flashes have
been made. Now it is widely believed that X-ray Flashes and Gamma-ray Bursts are
intrinsically the same, their much different peak energy and flux may be just
due to our different viewing angles to them. Here we analytically calculate the
very early afterglow of X-ray Flashes, i.e., the reverse shock emission powered
by the outflows interacting with the interstellar medium. Assuming $z\sim 0.3$,
we have shown that typically the R Band flux of reverse shock emission can be
bright to $16\sim 17$th magnitude (the actual values are model dependent and
sensitive to the initial Lorentz factor of the viewed ejecta). That emission is
bright enough to be detected by the telescope on work today such as ROTSE-III or
the upcoming UVOT carried by Swift Satellite, planned for launching in later
2004.

\end{abstract}

\begin{keywords}
X-rays: general---Gamma-rays: bursts---radiation mechanism:
non-thermal
\end{keywords}

\section{Introduction}

X-ray Flashes (XRFs) have received increasing attention in the
past several years (e.g., Heise et al. 2002; Kippen et al. 2002).
In many respects, XRFs are similar to ``classical'' Gamma-ray
Bursts (GRBs): (1) Their sky distribution is nearly isotropic; (2)
The red-shift $z=0.251$ of XRF 020903 (Soderberg et al. 2003) and
afterglows of XRF 020903 and XRF 030723 have been detected
(Soderberg et al. 2002; Prigozhin et al. 2003), which suggests
that XRFs are also cosmic events; (3) They have $T_{90}$ durations
ranging from 20 to 200 seconds; (4) Their spectrum can also be
fitted by the Band spectrum (Kippen et al. 2002); (5) They have
the temporal structure which is very similar to the X-ray
counterparts of GRBs (Heise et al. 2002). However, their spectral
peak energies $E_{\rm peak,obs}\sim 10 {\rm keV}$ are much lower
than that of GRBs ($\sim 300{\rm keV}$). These similarities lead
to the suggestion that XRFs are the extension of GRBs and X-ray
rich GRBs to even softer regime (Kippen et al. 2002; Lamb, Donaghy
\& Graziani  2004).

Now there are several models proposed to account for XRFs, i.e.,
the off-beam uniform jet model (e.g., Ioka \& Nakamura 2001;
Yamazaki, Ioka \& Nakamura 2002); the wide opening angle uniform
jet model (e.g., Lamb et al. 2004); the Gaussian jet model (e.g.,
Zhang \& M\'{e}sz\'{a}ros 2002a; Lloyd-Ronning et al. 2004; Zhang
et al. 2004a); the power-law jet model (e.g., M\'{e}sz\'{a}ros,
Rees \& Wijers 1998; Jin \& Wei 2004); the two component jet model
(e.g., Zhang, Woosley \& Heger 2004; Huang et al. 2004) and the
Cannon-ball model (Dado, Dar \& De Rugula 2003) .

In this Letter, instead of taking a further insight into these models, we turn
to calculate the possible accompanying very early afterglows. We are mostly
incited by the upcoming Swift
Satellite\footnote{http://swift.gsfc.nasa.gov/science/instruments}, which
carries three main telescopes---The Burst Alert Telescope (BAT), the X-ray
Telescope (XRT), the Ultraviolet and Optical Telescope (UVOT). The energy range
of BAT is 15-150 keV. Considering its high sensitivity, XRFs with not too much
lower peak energies can be detected as well as GRBs. BAT will observe and locate
hundreds of bursts per year to better than 4 arc minutes accuracy. Using this
prompt burst location information, Swift can slew quickly to point on-board XRT
and UVOT at the burst for continued afterglow studies. The spacecraft's 20$-$70
seconds time-to-target means that about $\sim$100 GRBs+XRFs per year (about 1/3
of the total) will be observed by the narrow field instruments during the gamma
ray emission. The UVOT is sensitive to magnitude 24 in a 1000 seconds exposure
(For a linear increasing of the sensitivity with the exposure time, that means a
sensitivity of magnitude 19 in a 10 seconds exposure). Then, though the very
early afterglow of XRFs may be dimmer than that of GRBs,  they can be detected
directly.\footnote{By now, there is only one upper limit ($m_{\rm R}>19$) at a
time $\sim 50{\rm s}$ after the main burst of XRF 030723 has been reported
(Smith et al. 2003)}

\section{The very early afterglow of GRBs}
In the standard fireball model for Gamma-ray bursts, the very
early afterglow of GRBs powered by the ejecta interacting with the
interstellar medium (ISM) or stellar wind has been discussed in
great detail (e.g., Sari \& Piran 1999; M\'{e}sz\'{a}ros \& Rees
1999; Kobayashi 2000; Li et al. 2003; Nakar \& Piran 2004; Wu et
al. 2003; Kobayashi \& Zhang 2003a). Recently, the reverse shock
(RS) emission powered by magnetized outflows---medium interaction
has been discussed by Fan, Wei \& Wang (2004) and Zhang \&
Kobayashi (2004) independently. So far there are three very early
afterglows having been detected (see Sari \& Piran 1999; Kobayashi
\& Zhang 2003b; Wei 2003 and the references therein). With the
launching of Swift, that number may be increased greatly.

Modelling the very early afterglow can be used to constrain some
poorly known physical parameters, such as the initial Lorentz
factor of the outflow, $\epsilon_{\rm B}$ (the fraction of the
internal energy converted into magnetic energy) and so on (e.g.,
Wang et al. 2000). Interestingly, it is found that the RS emission
regions of GRB 990123 and GRB 021211 are likely to be magnetized
(Fan et al. 2002; Zhang, Kobayashi \& M\'{e}sz\'{a}ros 2003; Kumar
\& Panaitescu 2003).

\section{The very early Afterglow of XRFs}
One of obstacles we encountered in the current work is the poorly
known angular distribution of the initial Lorentz factor of these
jets except the off-beam one. Kumar \& Granot (2003) have
performed a numerical investigation on the hydrodynamical
evolution of a Gaussian jet by assuming $\eta(\theta)$ is also
Gaussian distribution. However, if the viewed Lorentz factor
$\eta({\rm \theta_{\rm v}})$ is significantly lower than 100 and
XRFs are powered by internal shocks, the observed spectrum should
be thermal, which is inconsistent with the current observation. In
their Monte Carlo simulation, Zhang et al. (2004) have taken
$\eta(\theta)$ as a free function but found that a flat (or at
most slightly variable) angular distribution of Lorentz factor is
indeed required to interpret the current observations,
particularly, the empirical relationship\footnote{Its validity for
GRBs and XRFs has been verified by many authors (e.g.,
Lloyd-Ronning, Petrosian \& Mallozzi 2000; Amati et al. 2002;
Zhang \& M\'{e}sz\'{a}ros 2002b; Wei \& Gao 2003;  Lamb et al.
2004; Liang, Dai \& Wu 2004).} $E_{\rm peak}\propto E_{\rm
iso}^{1/2}$. Therefore, and partly for convenience, we assume
$\eta(\theta_{\rm v})={\rm const}\sim 150$. For the off-beam jet
model, XRFs are intrinsically the GRBs. As usual, we take
$\eta=300$.

In this work, we assume $\eta(\theta_{\rm v})={\rm const}\sim 150$
for all on-beam jets (i.e., the Gaussian jet, the power-law jet,
the wide opening jet and the two component jet).   Different from
the off-beam one, for these four type jets, there are
relativistically moving materials beaming towards the observer. As
long as the Lorentz factor is large enough, within the $1/\gamma$
cone the jet structure effect is not significant, all these four
models would give a rather similar early afterglow lightcurves.
Their differences would only appear in much later epochs when the
jet-structure effect becomes prominent. So, for an illustrative
purpose, in this Letter,  only the possible RS emission powered by
the Gaussian jet and the off-axis jet interacting with ISM have
been calculated.

\subsection{The RS emission of the off-beam jet}
For a uniform jet, if the line of sight is slightly beyond the
cone, i.e., $\Delta \theta \equiv \theta_{\rm v}-\theta_{\rm
jet}>0$ ($\theta_{\rm jet}$ is the opening angle of jet,
$\theta_{\rm v}$ is our viewing angle to the jet), the observed
peak energy decreases as
\begin{equation}
\nu_{\rm off}=a\nu_{\rm on},
\end{equation}
where $\nu_{\rm on/off}$ are the observed frequency on/off-axis respectively,
$a\approx [1+\gamma^2 (\Delta \theta)^2]^{-1}$ (e.g., Granot et al. 2002). On
the  contrary, the observed duration is increased by a factor of $a^{-1}$. Then
a ``classical'' GRB will be detected as a long lasting XRF (e.g., Ioka \&
Nakamura 2001; Yamazaki et al. 2002). By assuming an accompanying supernova, it
is claimed that the off-beam jet model can fit the later afterglow of XRF 030723
quite well (Fynbo et al. 2004).

The afterglow of off-beam jet has been numerical calculated by
many authors, and an empirical formula has been proposed to
estimate the observed flux (e.g., Granot et al. 2002; Jin \& Wei
2004)
\begin{equation}
F_{\rm \nu_{\rm off}}(\Delta\theta ,t_{\rm obs})\approx {a^3\over
2}F_{\nu_{\rm on}}(0,t).\label{Eq:Flux}
\end{equation}
where $dt_{\rm obs}=dt/a$. Different from the long lasting forward
shock (FS), the RS disappears after it crosses the ejecta. The
crossing time is estimated by $t_{\rm \times}=\max\{t_{\rm
dec},~T_{\rm dur}\}$, where two timescales are involved: (i) The
deceleration time $t_{\rm dec}$, which can be calculated as
follows: The outflow is decelerated significantly at the
deceleration radius (Rees \& M\'{e}sz\'{a}ros 1992)
\begin{equation}
R_{\rm dec}\approx 5.6\times 10^{16}~{\rm cm}~E_{\rm
iso,53}^{1/3}n_{\rm 1,0}^{-1/3}\eta_{2.5}^{-2/3},\label{Eq:Rdec}
\end{equation}
where $E_{\rm iso}\sim 10^{53}{\rm ergs}$ is the typical isotropic
energy of the GRB outflow, $n_1\sim 1$ is the number density of
the ISM, $\eta\sim 300$ is the initial Lorentz factor of the
outflow at the end of $\gamma-$ray emission phase. Throughout this
Letter, we adopt the convention $Q_{\rm x}=Q/10^{\rm x}$ for
expressing the physical parameters, using cgs units. At $R_{\rm
dec}$, the Lorentz factor of the outflow drops to $\gamma_{\rm
dec} \simeq \eta/2$. The corresponding timescale is
\begin{equation}
t_{\rm dec}\approx R_{\rm dec}/2\gamma_{\rm dec}^2 c=40~{\rm
s}~R_{\rm dec, 16.75}\gamma_{\rm dec,2.18}^{-2}.\label{Eq:tdec}
\end{equation}

(ii) $T_{\rm dur}$, the local duration of the GRB corresponding to
the XRF, can be estimated as follows: In the off-beam jet model of
XRFs, the observed duration of XRF is $\sim(1+z)a_0^{-1}T_{\rm
dur}$, where $a_0 \equiv [1+\eta^2(\Delta \theta)^2]^{-1}$. Here
we take $a_0 \simeq 0.03$---For much smaller $a_0$, the observed
luminosity ($L\propto a_0^4$) is too dim to be detected; for much
larger $a_0$, the observed peak energy ($\propto a_0$) is in the
hard X-ray energy, i.e., what we observed is X-ray rich burst
rather than XRF. One potential problem of the off-beam model is
that, in principle, the typical duration of XRFs should be tens
times of that of typical GRBs---which has not been supported by
the present observations (The observed duration of the XRFs
ranging from 20 s to 200 s). Here we take $T_{\rm dur}\sim 10{\rm
s}$, which matches that of the ``classical'' long GRBs ($\sim
20{\rm s}/(1+z')$, $z'\sim 1$ is the typical red-shift of GRBs).

For $t_{\rm dec}>T_{\rm dur}$, the shell is thin, otherwise the
shell is thick (Sari \& Piran 1999; Kobayashi 2000). For the
parameters taken here, we have $t_{\rm dec}>T_{\rm dur}$, so the
shell is thin (If the shell is thick, the following discussion is
invalid and we refer the reader to see $\S3.2$ for detailed
treatment). Consequently, $t_{\rm \times}=t_{\rm dec}$,
$\gamma_{\rm \times}=\gamma_{\rm dec}$ and $R_{\rm \times}=R_{\rm
dec}$ ($\gamma_{\rm \times}$ is the bulk Lorentz factor of the
outflow at $t_{\times}$, $R_{\rm \times}$ is the corresponding
radius).

The Lorentz factor of the shocked outflow relative to the initial
one is
\begin{equation}
\gamma_{\rm 34,\times}\approx (\eta/\gamma_{\rm
\times}+\gamma_{\rm \times }/\eta)/2=1.25,
\end{equation}
which suggests that the  RS is only mildly relativistic.

At $R_{\rm \times}$, all the electrons contained in the outflow
have been heated by the  RS and distribute as $dn/d\gamma_{\rm
e}\propto \gamma_{\rm e}^{\rm -p}$ for $\gamma_{\rm e}>\gamma_{\rm
e,m}$ (Throughout this Letter we take $p=2.2$), where the
``minimal'' thermal Lorentz factor, $\gamma_{\rm e,m}$, can be
estimated by
\begin{eqnarray}
\gamma_{\rm e, m}={m_{\rm p}\over m_{\rm e}}{\rm p-2\over
p-1}\epsilon_{\rm e}(\gamma_{\rm 34,\times}-1)=23F,
\end{eqnarray}
where $F\equiv \epsilon_{\rm e,-0.5}({\gamma_{34,\times}-1\over
0.25})$, $\epsilon_{\rm e}$ is the fraction of thermal energy
obtained by the electrons, $m_{\rm p}~(m_{\rm e})$ are the rest
mass of proton (electron) respectively. The typical synchrotron
radiation frequency can be estimated by
\begin{eqnarray}
\nu_{\rm m,\times}&=&{\gamma_{\rm e,m}^2\gamma_{\rm \times}eB\over
2(1+z)\pi
m_{\rm e}c}\nonumber\\
&=&{4.0\times 10^{12}\over 1+z}~{\rm Hz}~F^2\epsilon_{\rm
B,-1}^{1\over 2}n_{1,0}^{1\over 2}\gamma_{\rm \times,2.18}^2,
\end{eqnarray}
where $B\approx 0.12{\rm G}~n_{1,0}^{1/2}\epsilon_{\rm
B,-1}^{1/2}\gamma_{\rm \times}$ (e.g., Fan et al. 2002) is the
magnetic strength generated in the  RS. Here we take
$\epsilon_{\rm B}\sim 0.1$ rather than 0.01 since the ejecta is
likely to be magnetized (e.g., Fan et al. 2002; Zhang et al.
2003). The cooling Lorentz factor of the shocked electrons is
(Sari, Piran \& Narayan 1998, hereafter SPN) $\gamma_{\rm
c,\times}\approx 6\pi m_{\rm e}c/(\sigma_{\rm T}\gamma_{\rm
\times}B^2 t_{\rm \times})\approx 360$ (Assuming the involved
$Q_{\rm x}=1$), where $\sigma_{\rm T}$ is the Thompson cross
section. Correspondingly, the cooling frequency is $\nu_{\rm
c,\times}= ({\gamma_{\rm c,\times}\over \gamma_{\rm e,m}})^2
\nu_{\rm m}\approx 10^{15}/(1+z)~{\rm Hz}$. For $z=0.3$ (as it is
general suggested for XRFs), $\nu_{\rm c,\times}$ is larger than
the observer frequency $\nu_{\rm R,obs}=4.6\times 10^{14}~{\rm
Hz}$. Following Wu et al. (2003; see their appendix A1 for
detail), the synchrotron self-absorption frequency of the
compressed outflow is $\nu_{\rm a,\times}\approx 3.6\times
10^{12}/(1+z)~{\rm Hz}$ (Assuming the involved $Q_{\rm x}=1$). So
the synchrotron self-absorption effect can not significantly
change the R band spectrum, which is of our interest.

At $R_{\rm \times}$, the on-beam peak flux can be estimated by
$F_{\rm \nu, max(\rm on)}\approx (1+z)N_e m_{\rm e}c^2\sigma_{\rm
T}\gamma_{\rm \times}B/ 12\pi e D_{\rm L}^2\approx 92.3({1+z\over
1.3})~{\rm Jy}~E_{\rm iso,53}\eta_{2.5}^{-1}\gamma_{\rm
\times,2.18}^2n_{1,0}^{1/2}\epsilon_{\rm B,-1}^{1/2}D_{\rm
L,27.7}^{-2}$ (SPN), where $N_{\rm e}=E_{\rm iso}/\eta m_{\rm
p}c^2$ is the total number of electrons contained in the outflow,
$D_{\rm L}$ is the luminosity distance (we assume $H_0=70 \rm km$
$\rm s^{-1}$ $\rm Mpc^{-1}$, $\Omega_{\rm M}=0.27$, $\Omega_{\rm
\wedge}=0.73$). With equation (2), the off-axis observed energy
flux can be estimated by (SPN)
\begin{eqnarray}
F_{\rm \nu_{\rm R,obs}(off)}&\approx & {a_{\rm \times}^3\over
2}F_{\rm \nu,max(on)}({\nu_{\rm
R,obs}\over a_{\rm \times}\nu_{\rm m,\times}})^{\rm -(p-1)/2}\nonumber\\
&=&0.8{\rm mJy}~({9a_{\rm \times}})^{\rm (p+5)\over 2}E_{\rm
iso,53}\eta_{2.5}^{-1}\gamma_{\rm \times,2.18}^{\rm p+1}
\nonumber\\
&&F^{\rm p-1} \epsilon_{\rm B,-1}^{\rm p+1\over 4}n_{1,0}^{\rm p+
1\over 4}({1+z\over 1.3})^{\rm 3-p\over2}D_{\rm L, 27.7}^{-2},
\end{eqnarray}
where $a_{\rm \times}=[1+(\gamma_{\rm \times}\Delta
\theta)^2]^{-1}$. If we take $a_0=0.03$, i.e., $\eta \Delta
\theta=5.7$ (see the reasons mentioned in the paragraph below
equation ({\ref{Eq:tdec}})), we have $a_{\rm \times}=1/9$, i.e.,
$\gamma_{\rm \times}\Delta \theta=2.8$. The observed crossing time
$t_{\rm \times,obs}\sim t_{\rm \times}/a_{\rm \times}\approx
360(1+z){\rm s}$. At R band, the magnitude $m_{\rm R}\approx 17$,
which is bright enough to be detected by the telescopes on work
today, or the upcoming UVOT.

Here, we briefly discuss the observed very early afterglow light curve. In the
case of thin shell, the on-axis light curve is well approximated by (for
$\nu_{\rm m}<\nu<\nu_{\rm c,\times}$, i.e., the slow cooling case)
\begin{equation}
 F_{\rm \nu(on)} \propto \left\{
\begin{array}{lll}
   t^{\rm 2p}, & {\rm for}\,\, t<t_{\rm \times} \sim t_{\rm dec}, \\\
   t^{-(\rm 11p+3)/14}, & {\rm for}\,\, t>t_{\rm \times} \sim t_{\rm dec}.
   \end{array} \right.\label{Eq:L1}
\end{equation}
The light curve for $t>t_{\rm \times}$ is presented in
M\'{e}sz\'{a}ros \& Rees (1999; Assuming the comoving magnetic
field contained in the shocked outflow is freezing and taking
$g=3$, since for the parameters adopted in this letter, the FS
emission is in fast cooling). The light curve for $t<t_{\rm
\times}$ adopted here is slightly different from that of Kobayashi
(2000) and Fan et al. (2002). Below we derive it in some detail.
For the Newtonian  RS, the bulk Lorentz factor of the ejecta
$\gamma\sim \eta$, $\gamma_{34}-1\propto (n_4/n_1)^{-1}\propto
R^2$ (Sari \& Piran 1995), where $n_4$ is the comoving number
density of the outflow, $R\simeq 2\eta^2 c t$ is the radius of the
outflow. Therefore $\gamma_{34}-1\propto t^2$, substituting it
into equation (7) we have $\nu_{\rm m}\propto t^4$. On the other
hand, $\beta_{34}\propto t$, which results in $N'_{\rm e} \propto
t^2$, where $N'_{\rm e}$ is the number of shocked electrons.
Substituting these relations into the expression of $F_{\rm
\nu,max(on)}$, we have $F_{\rm {\nu,max}(on)}\propto t^2$.
Therefore the observed R band light curve $F_{\rm \nu_{\rm R
,obs}(on)}\propto t^2 (t^{-4})^{-(p-1)/2}\propto t^{\rm 2p}$.

Assuming $\gamma=\eta(1-t/2t_{\times})$ for $t<t_{\rm \times}$, we
have $t_{\rm obs}\approx [{1\over a_0}-{1\over 2}({1\over
a_0}-1){t\over t_{\rm \times}}]t$; For $t>t_{\rm \times}$,
$\gamma=\eta (t/t_{\rm \times})^{-3/7}/2$, we have $t_{\rm
obs}\approx {5(a_0-1)\over 4a_0}t_{\times}+[1+{7\over 4
}({1-a_0\over a_0})({t_{\times}\over t})^{6/7}]t$. With these
relations (including eqs. (\ref{Eq:Flux}) and (\ref{Eq:L1})), we
obtain one sample light curve of the  RS emission powered by the
off-beam uniform jet--ISM interaction, which has been presented in
figure 1 (the solid line). Naturally, for $t_{\rm obs}<t_{\rm
obs,\times}$, the increasing of the light curve is more rapidly
than $t_{\rm obs}^{\rm 2p}$. For $t_{\rm obs}>t_{\rm obs,
\times}$, as long as $\gamma \Delta \theta$ is not much smaller
than 1,  the factor $a^{\rm (p+5)/2}$ increases rapidly.
Therefore, at early time, the optical emission increases, rather
than decreases with time. At much later time, $\gamma \Delta
\theta\rightarrow 1$, the factor $a^{\rm (p+5)/2}$ increases only
slightly. So the light curve of RS emission drops as $\propto
t_{\rm obs}^{-2}$ (Figure 1, the thin solid line).

Taking $\epsilon_{\rm B,f}=0.01$, $\epsilon_{\rm
e,f}=\epsilon_{\rm e}$ (the subscript $f$ represents the forward
shock), we have: At $t_{\rm \times,obs}$, $F_{\rm \nu_{\rm
R,obs(off)},f}\approx 0.04{\rm mJy}$. For $t>t_{\rm \times,obs}$,
$F_{\rm \nu_{\rm R,obs }(on),f}\propto t^{-1/3}$ (e.g., SPN). With
eq. (\ref{Eq:Flux}) and $t_{\rm obs}\approx {5(a_0-1)\over
4a_0}t_{\times}+[1+{7\over 4 }({1-a_0\over a_0})({t_{\times}\over
t})^{6/7}]t$, the sample early light curve of the FS emission has
been presented in Figure 1 (The dotted line).

\begin{figure}
\begin{picture}(20,180)
\put(0,0){\includegraphics{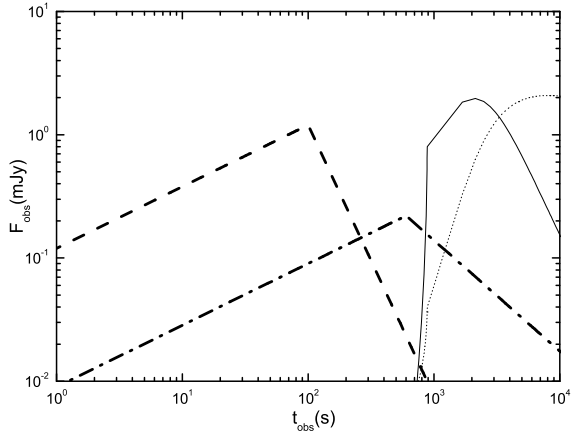}}
\end{picture}
\caption {The sample very early R-band ($\nu_{\rm R,obs}=4.6\times
10^{14}{\rm Hz}$) light curves powered by the outflows interacting
with the ISM. The solid/dotted lines represent the RS/FS emission
as a function of observer time of the off-beam uniform jet; the
thick dashed/dash-dotted lines represent the RS/FS emission as a
function of time of the Gaussian jet. The parameters for plotting
the solid/dotted lines are: $T_{\rm dur}=10{\rm s}$,
$\Delta\theta=0.019{\rm rad}$, $\eta=300$, $E_{\rm
iso}=10^{53}{\rm ergs}$, $z=0.3$, $p=2.2$, $\epsilon_{\rm
e}=\epsilon_{\rm e,f}=0.3$, $\epsilon_{\rm B}=0.1$, $\epsilon_{\rm
B,f}=0.01$. For plotting the thick dashed/dash-dotted lines, the
parameters are the same except $T_{\rm 90}=100{\rm s}$,
$\eta(\theta_{\rm v})=150$, $E_{\rm iso(\theta_{\rm
v})}=10^{50}{\rm ergs}$.}
\end{figure}

\subsection{The  RS emission of Gaussian jet}
For a Gaussian jet, the observed isotropic energy  $E_{\rm
iso(\theta_{\rm v})}=E_{\rm iso(\theta_{\rm v}=0)}
\exp({-\theta_{\rm v}^2/2\theta_0^2})$. Typically, the observed
peak energy of XRF is about 0.03 times that of GRBs and $E_{\rm
iso(\theta_{v})}\sim 10^{-3}E_{\rm iso(\theta_{\rm v}=0)}$. The
corresponding viewing angle is $\theta_{\rm v}\approx
3.7\theta_0$. Taking $\eta(\theta_{\rm v})=150$, equation (3)
gives $R_{\rm dec}\sim 8.8\times 10^{15}{\rm cm}$.
Correspondingly, $t_{\rm dec}\sim 25~{\rm s}$, which is much
shorter than the typical duration of the XRFs $T_{\rm 90}\sim
100~{\rm s}$, i.e., \textsl{the shell is thick}. So, locally
$t_{\rm \times}\approx T_{90}/(1+z)$, with which $R_{\rm \times}$
and $\gamma_{\rm \times}$ can be calculated self-consistently.

At $R_{\rm \times}$, the energy conservation of the system, i.e., the shocked
ISM and the shocked viewing outflow, gives
\begin{equation}
\gamma_{\rm \times}\gamma_{\rm 34,\times}M_{\rm ej}+\gamma_{\rm
\times}^2M_{\rm sw}\approx \eta(\theta_{\rm v}) M_{\rm ej},
\end{equation}
where $M_{\rm ej}$ ($M_{\rm sw}$) is the mass of the viewed ejecta
(the swept ISM). In the thick shell case, the  RS is
mild-relativistic and $\gamma_{34,\times}\approx \eta(\theta_{\rm
v})/2\gamma_{\rm \times}$. Now equation (10) reduces to
$\gamma_{\rm \times}^2M_{\rm sw}\approx \eta(\theta_{\rm v})
M_{\rm ej}/2$. Considering that $M_{\rm sw}={4\pi\over 3}R_{\rm
\times}^3n_1m_{\rm p}$ and $T_{90}/(1+z)\approx R_{\rm
\times}/2\gamma_{\rm \times}^2c$, we have
\begin{equation}
R_{\rm \times}\approx 1.4\times 10^{16}~{\rm cm}~E_{\rm
iso(\theta_{\rm v}),50}^{1/4}
n_{1,0}^{-1/4}T_{90,2}^{1/4}({1.3\over 1+z})^{1/4},
\end{equation}
\begin{equation}
\gamma_{\rm \times}\approx 55E_{\rm iso(\theta_{\rm v}),50}^{1/8}
n_{1,0}^{-1/8}T_{90,2}^{-3/8}({1+z\over 1.3})^{3/8}.
\end{equation}
Now $\gamma_{34,\times}\approx 1.55$, which is mild-relativistic.
So the assumption made before is reasonable. Similar to section
3.1, the typical frequency of  RS emission can be estimated by
\begin{equation}
\nu_{\rm m,\times}={3.8\times 10^{12}\over 1+z}~{\rm
Hz}~F_1^2n_{1,0}^{1\over 2}\epsilon_{\rm B,-1}^{1\over
2}\gamma_{\rm \times,1.74}^2.
\end{equation}
where $F_1\equiv \epsilon_{\rm e,-0.5}({\gamma_{\rm
34,\times}-1\over 0.55})$. Similarly, taking $Q_{\rm x}=1$ and
$z=0.3$ we have $\nu_{\rm c,\times}\sim 2.9\times 10^{16}{\rm
Hz}$. Following Wu et al. (2003), the synchrotron self-absorption
frequency is $\nu_{\rm a,\times}\sim 4.5\times 10^{10}~{\rm Hz}$.
Therefore both of them can not affect the R band spectrum
significantly. The peak flux of  RS emission can be estimated by
$F_{\rm \nu, max}\approx 28{\rm mJy}~({1+z\over 1.3})E_{\rm {\rm
iso(\theta_{\rm v})},50}\eta_{2.18}^{-1}\gamma_{\rm
\times,1.74}^2n_{1,0}^{1/2}\epsilon_{\rm B,-1}^{1/2}D_{\rm
L,27.7}^{-2}$. Then the observed peak energy flux can be estimated
by (SPN)
\begin{eqnarray}
F_{\rm \nu_{\rm R, obs}}&\approx & F_{\rm \nu,max}({\nu_{\rm
R,obs}\over \nu_{\rm m,\times}})^{\rm -(p-1)/2}\nonumber\\
&=&1.2{\rm mJy}~E_{\rm iso(\theta_{\rm
v}),50}\eta_{2.18}^{-1}\gamma_{\rm \times,1.74}^{\rm p+1}
{F_1}^{\rm p-1} \epsilon_{\rm B,-1}^{\rm
p+1\over 4}n_{1,0}^{p+1\over 4}\nonumber\\
&&({1+z\over 1.3})^{\rm 3-p\over2}D_{\rm L, 27.7}^{-2},
\end{eqnarray}
the magnitude $m_{\rm R}\approx 16$, which is also bright enough
to be detected by the upcoming UVOT or the telescopes on work
today, as long as the response to the XRF is fast enough.
%\textbf{Someone may have noted that the peak flux predicted here
%is similar to that yielded by equation (8). That similarity can be
%understood as follows: For the Gaussian jet, $E_{\rm
%iso(\theta_{\rm v})}\sim 10^{50}{\rm ergs}$, as long as
%$\gamma\gg1$. But for the off-beam jet, with the deceleration of
%the outflow, the beaming factor $[1+(\gamma\Delta \theta)^2]^{-3}$
%increases rapidly.  At $R_{\rm dec}$, the viewed isotropic energy
%is $a_{\times}^3E_{\rm iso}\approx 1.4\times 10^{50}{\rm ergs}\sim
%E_{\rm iso(\theta_{\rm v})}$ (Assuming $\eta \Delta\theta =5.7$)!}

In the case of thick shell,  the very early light curve of a
standard fireball takes the form (For the parameters adopted in
this Letter, the reverse/FS emission are all in slow cooling, for
$t>t_{\rm \times}$).
\begin{equation}
 F_{\rm \nu_{\rm R,obs}} \propto \left\{
\begin{array}{lll}
   t^{\rm 1/2}, & {\rm for}\,\, t<t_{\rm \times} \sim T_{90}/(1+z), \\\
   t^{-3(\rm 5p+1)/16}, & {\rm for}\,\, t>t_{\rm \times} \sim T_{90}/(1+z).
   \end{array} \right.
\end{equation}

Thanks to the beaming effect, these scaling laws may be applied to
the current work as well. Here we simply take equation (15) to
plot the sample very early light curve powered by Gaussian
jet---ISM interaction (See figure 1, the thick dash line). The
corresponding light curve of the FS emission (see figure 1, the
thick dash-dotted line) is plotted as follows: Following SPN, at
$t_{\rm \times,obs}=(1+z)t_{\times}$, $F_{\rm \nu_{\rm
R,obs},f}\approx 0.09{\rm mJy}$. For $t<t_0$, $F_{\rm \nu_{\rm
R,obs} ,f}\propto t^{1/2}$ (where $t_0$ is determined by $\nu_{\rm
m,f}(t_0)=(1+z)\nu_{\rm R,obs}$); For $t>t_0$, $F_{\rm \nu_{\rm
R,obs} ,f}\propto t^{-0.9}$.

As shown in Figure 1, in many respects, i.e., the peak time and
the temporal behavior, the current light curve is much different
from that powered by the off-beam jet---ISM interaction, which may
help us to distinguish them. For example, the FS peak emission of
the off-beam jet is much brighter than that of Gaussian jet, which
is mainly due to: In the off-beam jet model, we take
$\Delta\theta=0.019{\rm rad}$. At several thousand seconds after
the main burst, the ejecta has been decelerated significantly
($\gamma<50$), as a result, the beaming effect is not important.
So the viewed isotropic energy of the ``off-beam'' ejecta is
nearly $E_{\rm iso}$, which is much larger than that of the
Gaussian jet model ($\sim E_{\rm iso(\theta_{\rm v})}$)---In which
$\theta_{\rm v}\simeq 0.2{\rm rad}$ (Zhang et al. 2004), the
emission powered by the central energetic ejecta can not be
observed at early time as long as $\gamma\gg 5$.

\section{Discussion and Conclusion}
In the past several years, XRFs have received more and more attentions, but
their nature remains unknown. Considering the similarity between XRFs and GRBs
on duration, temporal structure, spectrum and so on, these bursts may be the
same phenomenon. The different peak energy as well as the peak energy flux may
be just due to our different viewing angles to them. This viewpoint has been
supported by the recent detection of two afterglows and one red-shift of XRFs.
However, there are more than 6 models have been proposed to explain the XRFs.
Some of them (for example, the Gaussian jet model, the off-beam uniform jet
model) work well on explaining the current observation. Perhaps only the
detailed multi-wavelength afterglows modelling (including the very early
afterglow discussed in this Letter) can provide us a reliable identification on
them.

With two current leading models, in this Letter, the very early
 RS emission of XRFs have been analytically
investigated. As XRFs are much dimmer than GRBs, the predicted
very early R band afterglow of XRFs is dimmer than those of GRBs.
But some of them, if not all, are still bright enough to be
detected by ROTSE-III on work today or by the upcoming Swift
mission. The actual results are model dependent, which may in turn
provide us a chance to see which one is better, if the very early
afterglow of XRFs has been really detected. One thing should be
emphasized here is that the predicted R band flux is sensitive to
the initial Lorentz factor, i.e.,  $F_{\rm \nu_{\rm R,obs}}\propto
\eta(\theta_{\rm v})^{-1} \gamma_{\rm \times}^{\rm
p+1}(\gamma_{\rm 34,\times}-1)^{\rm p-1}$. So, if
$\eta(\theta_{\rm v})\sim {\rm tens}$ rather than 150 taken in
this Letter,  the resulting very early optical emission will be
much dimmer.

As realized by more and more ``GRB people'', modelling the very
early afterglow of GRBs can impose some stringent constraint on
the fundamental physical parameters of the outflow such as the
initial Lorentz factor of the ejecta (e.g., Sari \& Piran 1999;
Wang et al. 2000), or help to see whether the outflows are
magnetized or not (e.g., Fan et al. 2002; Zhang et al. 2003; Kumar
\& Panaitescu 2003; Fan et al. 2004b; Zhang \& Kobayashi 2004). In
addition, the  RS emission powered by the ejecta---stellar wind
interaction is much different from that powered by the
ejecta---ISM interaction (e.g., Wu et al. 2003; Kobayashi \& Zhang
2003a; Fan et al. 2004b). Therefore the very early afterglow
observation can provide us an independent chance to determine the
environment where the ``classical'' GRB was born in. In principle,
interstellar wind environment may be common, and the predicted
very early R band emission is very strong ($m_{\rm R }\sim 9$ or
even brighter). So far there are only three very early afterglow
having been reported. All of them can be well fitted by the
ejecta---ISM interaction model (e.g., Sari \& Piran 1999;
Kobayashi \& Zhang 2003b; Wei 2003). Therefore, in this Letter
only the ejecta---ISM interaction case has been investigated.
Fortunately, it is straightforward to extend our treatment to the
wind case. Anyway, the importance of modelling the very early
afterglow of GRBs can be applied to that of XRFs as well.

In this Letter, the possible pair loading during the
$\gamma$/X-ray burst phase has not been taken into account. At
least for the off-beam uniform jet model, during the initial
$\gamma-$ray emission phase, large amount of electrons/positrons
may be created. The annihilation time scale is long and most of
generated pairs can not be annihilated locally. These pairs will
be heated by the  RS, too (Li et al. 2003, and the references
therein). Without doubt, the  RS emission has been significantly
softened. But, as argued by Fan, Dai \& Lu (2004), the R band
emission may be not. This can be easily understood, if there are
$k$ times electrons/positrons of that associated with the baryons,
current $\nu_{\rm m}$ would be $\nu_{\rm m}/k^2$. On the other
hand, current $F_{\rm \nu, max}$ would be $kF_{\rm \nu, max}$.
Then, in the slow cooling case, current $F_{\rm \nu,obs}$ should
be $k^{\rm 2-p}F_{\rm \nu,obs}$. For $p\sim 2.2$, such dependence
is far from sensitive. In other models for XRFs, during the X-ray
emission phase, at least in the viewed area, the possible pair
loading process is unimportant and can be ignored safely.

%%%%%%%%%%%%%%%%%%%%%%%%%%%%%%%%%
\section*{Acknowledgments}
Y. Z. Fan thanks Bing Zhang and Z. P. Jin for kind help. We also thank T. Lu, Z.
G. Dai, X. Y. Wang and X. F. Wu for fruitful discussions. We would like to
appreciate the anonymous referee for her/his valuable and detailed comments that
enable us to improve the paper significantly. This work is supported by the
National Natural Science Foundation (grants 10225314 and 10233010), the National
973 Project on Fundamental Researches of China (NKBRSF G19990754).

\end{document}